\documentstyle[12pt]{article}
\title{String junction as a baryonic constituent }

\vspace{2cm}

\author{Yu.S.Kalashnikova\thanks{e-mail:yulia@vxitep.itep.ru},
   A.V.Nefediev\thanks{e-mail:nefediev@vxitep.itep.ru}
\\ Institute of
Theoretical and Experimental Physics\\ 117259, Moscow,Russia}\date{}
\newcommand{\be}{\begin{equation}} \newcommand{\ee}{\end{equation}}
\begin{document}
\maketitle

\begin{abstract}

We extend the model for QCD string with quarks to consider the
Mercedes Benz string configuration describing the three-quark
baryon. Under the assumption of adiabatic  separation of quark
and string junction motion we formulate and solve the classical
equation of motion for the junction.We dare to quantize the motion
of the  junction, and discuss the impact of these modes on the
baryon spectra.

\end{abstract}

The $qqq$ baryon, together with the $q\bar{q}$ meson, is usually
considered
in the framework of the constituent model as a simplest system with
zero
triality. It is clear, nevertheless, that in the QCD -motivated
approach the
only way to form gauge-invariant states with three quarks in the
fundamental
colour representation is to introduce the so-called string
junction as an
object that couples three quarks to make a colour singlet:
$$
\Psi(x_1 x_2 x_3, x_0) =\ \ \ \ \ \ \ \ \ \ \ \ \ \ \ \ \
 \ \ \ \ \ \ \ \
\ \ \ \ \ \ \ \ \ \ \ \ \
$$
\be
\ \ \ \ \ \ \ \ \ \ \ \ \ \ \ \ \ =\psi^{\alpha}(x_1)\psi^{\beta}
(x_2)\psi^
{\gamma}(x_3)\Phi^{\alpha'}_{\alpha} (x_0, x_1)
\Phi^{\beta'}_{\beta} (x_0, x_2) \Phi^{\gamma'}_{\gamma} (x_0, x_3)
\varepsilon_{\alpha' \beta' \gamma'}\;,
\ee
where parallel transporters $\Phi$,
\be
\Phi^{\alpha}_{\beta} (x,y) = (P \exp ig \int^x_y B_{\mu}^a t^a
dz_{\mu})^{\alpha}_{\beta}\;\;,
\ee
stand to assure non-local gauge invariance of the $qqq$ system
placed into
non-trivial QCD background $B_{\mu}^a$. From this point of view
it is obvious
that the baryon  should be treated as a simplest multiquark system
containing, in comparison to the $q\bar{q}$ system, a new object
with it's
own dynamics. Standard way to deal with the junction dynamics is
to assume
that it always moves in such a way that the sum of distances
between quarks
and junction is minimal [1,2]. This assumption is motivated by
non-relativistic approach, where confinement manifests itself
as linear
potential between constituents. On the other hand, the idea
that a junction
might be responsible for certain  kinds of baryonic excitations
was already
discussed in [2].

In the present letter we relax the above-mentioned assumption
on a junction
following submissively the quarks. Actually our baryon is going
to contain
four constituents rather than three. Our starting point is the
Vacuum
Background Correlators method [3]. In the framework of this
method the
hadronic Green functions can be constructed starting from the QCD
Lagrangian. The dynamics of interaction is defined by the averages
of Wilson
loop operators, and it may be shown that the area law asymptotic
for Wilson
loop average gives rise to the string-type interaction of the
constituents.
The model for the resulting straight-line $q\bar{q}$ string was
considered
in [4], and the hybrid mesonic excitations were analysed in [5].

To formulate the Hamiltonian approach to the problem one is to
consider the
Green function of a baryon,
\be
G(x_1 x_2 x_3 x_0, y_1 y_2 y_3 y_0) = <\Psi(x_1 x_2 x_3, x_0)
\mid \Psi(y_1
y_2 y_3, y_0)>_B\;\;,
\ee
where the brackets mean the averaging over background field
configurations.
To define the effective action the Feynman-Schwinger
representation [6] of a
Green function (3) is used (in the Euclidean space):
$$
G(x_1 x_2 x_3 x_0, y_1 y_2 y_3 y_0) = \int^{\infty}_0 ds_1
\int^{\infty}_0 ds_2 \int^{\infty}_0 ds_3 \int Dz_1 Dz_2
Dz_3\times
$$
 \be
  \times\exp(-{\cal{K}}) \cdot <W>_B
  \ee
   where
   $$ W =
 \varepsilon_{\alpha \beta \gamma} \Phi^{\alpha}_{\alpha'}(\Gamma_1)
 \Phi^{\beta}_{\beta'}(\Gamma_2) \Phi^{\gamma}_{\gamma'}(\Gamma_3)
 \varepsilon^{\alpha' \beta' \gamma'}
$$
$$
{\cal{K}} = \sum^3_{i=1}
 (m^2_{q_i} s_i + \frac{1}{4} \int^{s_i}_0 \dot{z}^2_i (\tau) d\tau)
\;\;,
$$
boundary conditions are $z_i(0) = y_i\;, \; z_i(s_i) = x_i$,  and
the contours $\Gamma_i$ in the Wilson loop operator run over the
trajectories of the quarks. Using  the relation \be 1 = \frac{1}{3!}
\varepsilon_{\alpha_1 \alpha_2 \alpha_3} \varepsilon^{\beta_1 \beta_2
\beta_3} \Phi_{\beta_1}^{\alpha_1}(\Gamma)
\Phi^{\alpha_2}_{\beta_2}(\Gamma)\Phi^{\alpha_3}_{\beta_3}(\Gamma)
\ee
(which is due to the unimodularity condition for the SU(3)),
where $\Gamma$
is an arbitrary open path (the same for all three $\Phi$'s)
connecting
points $x_0$ and $y_0$, we represent the Wilson loop in (4) as
$$
W=SpW_1\  SpW_2\  SpW_3 - Sp(W_1W_2)\  SpW_3 -
 $$
\be
 -Sp(W_3\ W_1)\  SpW_2 - Sp(W_2\ W_3)\  Sp W_1 +
\ee
$$
  +Sp(W_1\  W_2\  W_3) + Sp(W_3\  W_2\  W_1)
$$
where $(W_i)^{\alpha}_{\beta}$ is the ordered exponent along
the contour
formed by the paths of $i$-th quark and $\Gamma$.\footnote[1]{In what
follows $\Gamma$ will be interpreted as the path of the junction.}

The averaging over background was done using the cluster expansion
method [3], generalized for the case of more than one Wilson loop
as well as for a loop with self intersection in [7]. The main
result of
this method is that under assumption of existence of  finite gluonic
correlation length $T_g$ the generalized area law asymptotics may be
obtained. For the Wilson loop (6) we write \be <W>_B = \int Dz_0
\exp(-\sigma S_1 - \sigma S_2 - \sigma S_3)\;\;, \ee where $\sigma$
is the
string tension and $S_i$ is the minimal area inside the contour
$C_i$, and
we integrate in (7) over the junction trajectories $\{ z_0\}$ in
accordance
with our intention to treat a junction as a degree of freedom.
The area law
(7) is held for the contours with average size much larger than
$T_g$, and
is violated only when the contours are nearly embedded into the same
plane.In what follows we neglect such special configurations.

We would like to note here, that the standard approach to the
junction motion
corresponds to taking the classical trajectory ${z_0}(\tau)$ in the
integral (7) or, equivalently, assuming $\sum^3_{i=1} S_i = min$.The
latter
condition can be reduced in the potential model to the assumption
that the
sum of distances between quarks and junction is minimal. In
the "string-type"
language the representation (7) means that we are interested
in the special
kinds of string excitations, which are absent for simple
$q\bar{q}$ string
configurations, and reveal themselves in multiquark systems.

Using the parametrization
$$
z_{i\mu} = (\tau, \vec{r}_i),\;\; z_{0\mu} = (\tau, \vec{r}_0)
$$
and introducing new dynamical variables $$ \mu_i(\tau) =
\frac{T}{2s_i}
\\\dot{z}_{i0} (\tau)\;\;\; 0\leq \tau \leq T\;\;, $$ we rewrite
the Green
function (4) as \be G = \int D\vec{r}_1 D\vec{r}_2 D\vec{r}_3 D
\vec{r}_0
D{\mu_1} D{\mu_2} D{\mu_3} \exp(-A) \ee with the effective action
\be A =
\int^T_0 d\tau \Biggl[\sum^3_{i=1} \biggl(\frac{m_i^2}{2\mu_i}
+ \frac{\mu_i
\dot{r}_i^2}{2} +\frac{\mu_i}{2}+ \sigma \int^1_0 d\beta_i
\sqrt{\dot{w}^2_i
w'^{2}_i -(\dot{w}_i w'_i)^2}\biggr)\Biggr]
\ee
where the surfaces $S_i$ are parametrized by the coordinates
$w_{i\mu}$,
and $\dot{w}_{i\mu}=\frac{\partial w_{i\mu}}{\partial\tau}\;,
\;w'_{i\mu}
=\frac{\partial w'_{i\mu}}{\partial\beta}$. This procedure
of reduction
the four-dimensional dynamics to the three-dimensional one (9)
was first
suggested in [4] to analyse the $q\bar{q}$ system. With
the straight-line
ansatz for the  minimal surfaces

$$
w_{i\mu}=z_{i\mu}(\tau)(1-\beta_i)+z_{0\mu}(\tau)\beta_i,
$$
we obtain the Lagrangian
\be
L=\sum_{i=1}^3\Biggl(\frac{m_i^2}{2\mu_i}+
\frac{m_i\dot r^2_i}{2}+\frac{\mu_i}{2}+
\sigma\rho_i\int^1_0 d\beta_i\sqrt{1+l_i^2}\Biggr),
\ee
$$
\vec l_i=\frac{1}{\rho_i}[\vec{\rho}_i\times ((1-\beta_i)\dot{\vec
r}_i+\beta_i \dot{\vec r}_0)]\;,~~ \vec{\rho_i}=\vec r_i-\vec r_0
$$
The Lagrangian (10) with the additional requirement
$\sum_{i=1}^3\rho_i=min$ was discussed in detail in [8] with
neglecting the
angular velocities $\vec l_i$ under the square roots (string
correction), and the attempt to account the string correction was
undertaken in [9].

Here we concentrate on the dynamics of junction, so we assume the
quarks to be heavy enough $(m_i\gg \sqrt{\sigma})$ to allow the
adiabatic treatment of the problem. Following again the proceduce of
[4] we introduce the  auxiliary fields $\nu_i,\eta_i$:
\be
G=\int\prod^3_{i=1}D\nu_i\prod^3_{i=1}D\eta_i\prod^3_{i=1}D\mu_i
D\vec r_1
D\vec r_2
D\vec r_3
D\vec r_0
exp(-A[\{\nu\},\{\eta\}]),
\ee
$$
A[\{\nu\},\{\eta\}]= \int^T_0 d\tau
\sum^3_{i=1}\Biggl(\frac{m_i^2}{2\mu_i}
+\frac{\mu_i\dot r_i^2}{2}+\frac{\mu_i}{2}+\int^1_0
d\beta_i\frac{\sigma^2(\vec{r}_i-\vec{r}_0)^2}{2\nu_i}+
$$
$$
+\frac{1}{2}\int^1_0 d\beta_i\nu_i\eta_i^2(\vec r_i-\vec r_0)^2+
\int^1_0 d\beta_i\nu_i\eta_i(\vec r_i-\vec
r_0,\beta_i\dot{\vec{r}}_0+(1-\beta_i) \dot{\vec r}_i)\Biggr)
$$

The integration over the fields $\nu_i$ and $ \eta_i$ actually
amounts to taking extremal values in $\nu_i$ and $\eta_i$.

To make the algebra as simple as possible we consider the case of
equal quark masses. Moreover, we take into account only the lowest
symmetric hyperspherical harmonic [10] in the quark subsystem, an
approximation which is proved to be very successful in few--body
systems. It is quite clear that under all these assumptions one has
$$
\mu_1=\mu_2=\mu_3=\mu,
\nu_1=\nu_2=\nu_3=\nu,
\eta_1=\eta_2=\eta_3=\eta,
$$
and, after integrating over the fields $\mu$ and $\eta$ we arrive
to the relatively simple Hamiltonian in the center--of--mass frame
(in the Minkowski space):
\be
H=3m+\frac{p^2+\frac{3}{4}q^2}{2m}+H_j
\ee
\be
H_j=\frac{Q^2}{2m_s}+\frac{r^2}{\rho^2}\frac{(\vec Q\vec
n)^2}{2m_s}+\frac{1}{2} m_s \omega^2 (r^2+\rho^2)+\frac{3}{2}\int^1_0
d\beta \nu,
\ee
$$
m_s=\int^1_0 d\beta\nu \beta^2,~~ m_s\omega^2=\sigma^2 \int^1_0
\frac{d\beta}{\nu},
$$
where the quark wave finction depends only on the hyperradius
$\rho,~~\rho^2=(\vec r_1-\vec r_2)^2+(\vec r_3- \frac{\vec r_1+\vec
r_2}{2})^2,$~~$ \vec r \equiv \vec r_0,
$ $ \vec n=\frac{\vec r}{r}$, and $\vec p, \vec q$ and $ \vec Q$ are
the momenta conjugated to the Jacobi coordinates
$\vec r_1-\vec r_2,~~\vec r_3-\frac{\vec r_1+\vec r_2}{2} $
and $\vec r$
correspondingly.

The adiabatic approximation means that one should find the
eigenenergies $E_n$ of the "fast" subsystem (13) as functions of
$\rho$, and  after that substitute $E_n(\rho)$ as adiabatic
potentials into the Hamiltonian (12) of "slow" subsystem.To perform
this program we first found the classical equations of motion
for the
Hamiltonian (13) with the result ( we restrict ourselves with
the spherically symmetric case of $L=0)$:

\be
E(a)=\frac{1}{2} m_s \omega^2\rho^2(a^2+1)+\frac{3}{2}\int^1_0
d\beta \nu, \ee
\be
(\vec Q\vec n)= m\omega
\rho\sqrt{\frac{a^2\rho^2-r^2}{\rho^2+r^2}},
\ee
where $a$ is the constant of integration.

It may be shown   from equations (14),
(15) that "freezing" of the
junction degree of freedom actually means substitution the lowest
classical energy $E(0)$  into the Hamiltonian (12). In this case
integration over $\nu$ is trivial, and one arrives to the potential
model Hamiltonian [1]\footnote[2]{The coefficient
at the confinement form in (16) differs slightly from one obtained
by taking lowest hyperspherical harmonic of the interaction with
$\sigma\sum_{i=1}^3\rho_i=min$. It happens because we integrate
over the field $\nu$ after hyperspherical decomposition rather than
before. The order of these procedures does not matter only if one
does not truncate the basis.}
\be
H=3m+\frac{p^2+\frac{3}{4}q^2}{3m}+\sqrt{3} \sigma\rho
\ee
On the  contrary, the Bohr--Zommerfeld quantization procedure for
the classical solution (14),(15) yields
\be
E_n=\frac{1}{2} m_s \omega^2\rho^2 (a^2_n+1)+\frac{3}{2}
\int^1_0d\beta \nu,
\ee
\be
m_s\omega \rho^2 J(a_n)=(2n+\frac{3}{2})\frac{\pi}{2},
\ee
where
\be
J(a)=\int^a_0 dx \sqrt{\frac{a^2-x^2}{1+x^2}},
\ee
and may be expressed  in terms of elliptic integrals
${\bf K}(\frac{a}{\sqrt{1+a^2}})$ and ${\bf E}
(\frac{a}{\sqrt{1+a^2}})$. The
integration over $\nu$ now cannot  be performed analytically, apart
from the case of asymptotically large $\rho$, $\rho\gg
\frac{1}{\sqrt{
\sigma}}$, where  the spectrum (17) becomes the spectrum of
harmonic oscillator, and the adiabatic potential takes the form
\be
E_n(\rho)=\sqrt{3}\sigma\rho+(2n+\frac{3}{2})\cdot \frac{3}{\rho}.
\ee
We can find, however, the value of $\rho$ that corresponds to the
minimum of the potential well for  quarks:
\be
\rho^2_0=\frac{\pi(2n+\frac{3}{2})}{2m_s\omega^2J(a^{(0)})},
\ee
where for any n $a^{(0)}\approx 2.2$ is the root of the equation
\be
2aJ(a)-(a^2+1)\frac{\partial J(a)}{\partial a}=0
\ee
Performing the integration over $\nu$ for $\rho_0$ defined from the
equation (21), we arrive to
\be
\nu_{ext}=\sqrt{\frac{\sigma}{3}(2n+\frac{3}{2})
\frac{{a^{(0)}}^2+1}{J(a^{(0)})}}(1-\beta^2)^{-1/2},
\ee
$$
\rho^2_0=\frac{4(n+\frac{3}{4})}{\sigma J (a^{(0)})},~~
E_n=\sqrt{\frac{3\pi^2\sigma}{4}(2n+\frac{3}{2})
\frac{{a^{(0)}}^2+1}{J(a^{(0)})}}
$$
Having in mind that our quarks are heavy( $m\gg \sqrt{\sigma}$) one
should expand the adiabatic  potential $E_n(\rho)$ (17) around
$\rho_0$ given by (23) using the extremal $\nu_{ext}$ up to the terms
$\sim (\rho-\rho_0)^2$. In such a way each adiabatic potential
$E_n(\rho)$ gives rise to  the whole family of excitations in the
quark subsystem.

Strictly speaking, the adiabatic potentials (17) are valid either
for $n\gg 1$, or for $\rho\gg\frac{1}{\sqrt{\sigma}}$, where
the spectrum
(20) is exact. It is known, however, that quasiclassical results
are usually rather accurate for small excitation numbers.   Using
with a lot of cautions the equations (22), (23) for the ground
state energy $E_0(\rho)$, we arrive to the effective one-dimensional
adiabatic potential:
\be
V(\rho)=\frac{{\cal{L}}^2}{2m\rho^2}+E_0(\rho),
\ee
where ${\cal{L}}^2=15/4$ defines the "hypercentrifugal" barrier
for the
lowest hyperspherical harmonic [10]. In accordance with  equations
(17), (18) adiabatic potentials diverge at small $\rho$ , or at
large $a^2_n$, but this divergence is  only the logarithmic one, as
it is easily seen from  the asymptotical behaviour of $J(a)$ (see
equation (19)). So for "realistic" values of quark masses
hypercentrifugal barrier dominates the effective potential (24) at
small values of $\rho$. Nevertheless, the spectrum of the ground
state family differs substantially from the one given by the
potential model (16), and it happens due to the account of
junction zero oscillations.

The confining force is not the whole story, and should be supplied
by the Couloumb pair--wise interaction at small distances. Moreover,
at small $\rho$ the area law (7) is strongly contaminated by the
effects of finite correlation length $T_g$, so that the string
regime does not develop itself at full scale. The explicit
calculations which take into account both these effects are in
progress now, and will be reported elsewhere. It would be rather
interesting to study the $qqq$ system of light quarks in the
presented approach, but this problem is much more formidable
because the adiabatic approximation is not applicable for the light
quarks.

We acknowledge the extremely useful discussions with K.G.Boreskov,
O.V.Kancheli and Yu.A.Simonov. This research is supported by the
Russian Fundamental Research  Foundation, Grant  No 93-02-14937,
and by INTAS - 93-0079.

    \end{document}